# No stagnation in the growth of population

Ron W Nielsen aka Jan Nurzynski[1]

Environmental Futures Centre, Gold Coast Campus, Griffith University, Qld, 4222, Australia

November, 2013

Growth of human population shows no signs of stagnation. The only small disturbance is identified as being probably associated with the coinciding impacts of *five* demographic catastrophes. The concept of the Epoch of Malthusian Stagnation is convincingly contradicted by empirical evidence.

**Introduction**[2]

In the first publication (Nielsen aka Nurzynski, 2013a) of this series of four we have discussed the concept of the Epoch of Malthusian Stagnation, the first stage of growth proposed by the Demographic Transition Theory. In the next publication (Nielsen aka Nurzynski, 2013b) we have analysed the data describing the growth of population in Sweden and the data for the correlation between the intensity of positive checks and the annual growth rate. We have drawn a series of conclusions about the mechanism of growth of human population, one of them that positive checks do not suppress growth but stimulate it by activating the efficient Malthusian regeneration mechanism (Malthus, 1798; Nielsen aka

---

[1] r.nielsen@griffith.edu.au; ronwnielsen@gmail.com; http://home.iprimus.com.au/nielsens/ronnielsen.html

[2] Discussion presented in this publication is a part of a broader study, which is going to be described in a book (under preparation): *Population growth and economic progress explained.*

Nurzynski, 2013a). We have pointed out that empirical evidence strongly suggests that the mechanism of Malthusian stagnation never worked and that the Epoch of Malthusian Stagnation did not exist. In the same publication we have concluded (predicted) that demographic catastrophes in the past were too weak to have any tangible effect on the growth of human population. This prediction was confirmed in the third publication (Nielsen aka Nurzynski, 2013c) supporting our earlier conclusion that the Epoch of Malthusian Stagnation did not exist. The final test now is to examine the time-dependent distribution of the growth of human population during the AD era to see whether they any suggestion of stagnant state of growth. More information on this topic will be presented in the forthcoming book mentioned in the footnote to this *Introduction*.

**World population data**

World population data (Manning, 2008; US Census Bureau, 2013 and references therein) between AD 1 and 1800, covering the AD part of the mythical Epoch of Malthusian Stagnation are presented in Fig. 1. Procedures adopted in estimating historical populations are discussed extensively by Durand (1967) and by Caldwell and Schindlmayr (2002). Listed in this figure and in Table 1, are the most prominent and the most lethal demographic catastrophes discussed in the previous publication (Nielsen aka Nurzynski, 2013c).

**Table 1**. The most significant demographic catastrophes with their estimated maximum death tolls.

| Event | Time (AD) | Death Toll |
|---|---:|---:|
| Red Eyebrows Revolt | 2-88 | 29,000,000 |
| Plague of Justinian | 541-542 | 25,000,000 |
| An Lu-Shan Rebellion | 756-763 | 36,000,000 |
| Mongolian Conquest | 1260-1295 | 40,000,000 |
| Great European Famine | 1315-1318 | 7,500,000 |
| Famine in China | 1333-1348 | 9,000,000 |

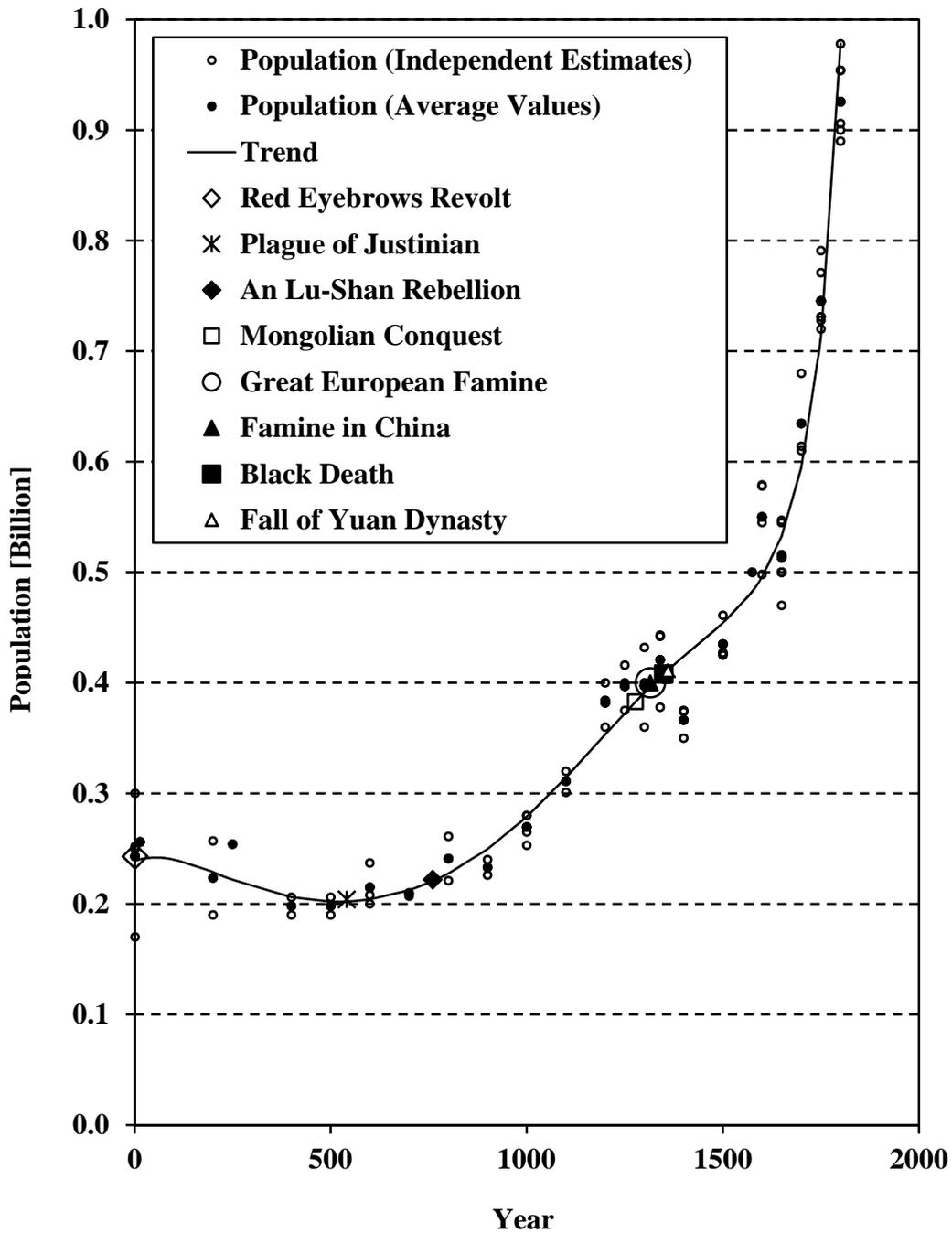

**Fig. 1.** World population data (Manning, 2008; US Census Bureau, 2013) between AD 1 and 1800 show a remarkably stable growth of the world population. Empirical evidence is again in direct contradiction with the legend of the Epoch of Malthusian Stagnation.

Fig. 1 contains interesting and important information:

1. It shows that there is a remarkably close agreement between the independent estimates of the size of human population. Indeed, in general, they are within ±10% of their corresponding averaged values. The largest deviations of around ±30% are for the AD 1 data. The two estimates for AD 200 differ by ±15% from their average value.

2. The estimated values follow a remarkably stable trajectory. There is nothing in them to suggest any form of prolonged stagnation. The claim that "cyclical behavior in *population growth* is often encountered in historical data" (Lagerlöf, 2006, p. 118, italics added) and all other similar claims of unstable and chaotic state of growth of human population discussed extensively in our first publication (Nielsen aka Nurzynski, 2013a) are contradicted by data.

3. The data contradict also the frequently repeated claim that demographic catastrophes were strongly influencing the growth of human population. The generally small damage inflicted by them (Nielsen aka Nurzynski, 2013c) must have been quickly repaired by the Malthusian regeneration mechanism (Malthus, 1798; Nielsen aka Nurzynski, 2013b).

4. The data show that at least during the displayed time *the Epoch of Malthusian Stagnation did not exist.* However, the analysis extending down to 300,000 BC, discussed in the forthcoming book, shows that the Epoch of Malthusian Stagnation never existed.

5. The only distortion in the growth trajectory displayed in Fig. 1 was possibly caused by *a cluster of five* closely-spaced demographic catastrophes: Mongolian Conquest, Great European Famine, the 15-year Famine in China commencing in 1333, Black

Death and the Fall of Yuan Dynasty, killing the combined total of 139 million people within a relatively short time, the most lethal combination ever recorded, but even these events, these exceptional and remarkably strong concentration of lethal forces, had only minor effect on the growth of population, showing that in contradiction with the concept of the Epoch of Malthusian Stagnation, the mythical first stage of growth claimed by the Demographic Transition Theory, the growth of human population was strong and stable.

**Two ways of processing information**

The difference between scientific and unscientific processing of information based on observations is summarised in Table 2 using as an example the growth of the human population.

Table 2. Two ways of processing information based on observations

| Method | Observation | Processing of information | Conclusions |
| --- | --- | --- | --- |
| Scientific | Over a long time, the size of human population was small. | **Rigorous analysis:** Population growth was following a well-defined trajectory. | **Evidence-based conclusions advancing knowledge:** Population growth is likely to have been prompted by a certain strong driving force, whose basis characteristics remained unchanged over a long time. |
| Unscientific | Over a long time, the size of human population was small. | **Imagination, impressions:** Population growth was stagnant. | **Stories, legends and myths:** Population growth was prompted by many random forces creating the Epoch of Malthusian Stagnation. Population was locked in the Malthusian trap. Many other stories and explanations discussed earlier (Nielsen aka Nurzynski, 2013a). |

The scientific and unscientific cognition processes illustrated by this example has the same starting point: the observation. However, while scientific approach is to analyse rigorously observed phenomena, the unscientific approach will prefer to take an easy way out and interpret them using a good dose of imagination based on impressions and beliefs. Impressions may be so strong that they can deceive even a strong intellectual. "It is clear that the earth does not move, and that it does not lie elsewhere than at the centre" declared Aristotle.

The next step is to offer an explanation of the observed phenomenon. Scientific approach will consist in a cautious formulation of conclusions based firmly on the analysis of empirical evidence. Unscientific approach will consists in the often unrestricted use of creative imagination leading to confidently proclaimed claims and descriptions, which would be hard or even impossible to verify by empirical evidence.

Conclusions derived using scientific approach might not be always accurate but they will be always useful because by being based on a careful analysis of available empirical evidence they are likely to point in the right direction to conduct further research. Conclusions formulated using unscientific approach might be interesting, attractive, appealing and even perhaps spectacular but they are unreliable and misleading. They are frequently based firmly on preconceived ideas, which often prompt to ignore empirical evidence rather than to use it; ideas inspired by personally experienced impressions or by the words of wisdom passed on by authorities, whose insights and inspirations are not supposed to be questioned. The real harm is not in just having two diametrically different approaches to the processing of information and to the interpretations of observed phenomena but in accepting unscientific explanations, claims and conclusions in the scientific literature because by doing so they receive a strong stamp of approval and they lead away from the correct line of the scientific investigation.

**Summary and conclusion**

The study presented here is the last in the series of four. This study reinforces our earlier conclusion that the concept of the Epoch of Malthusian Stagnation is unsupported by the empirical evidence and it suggests an entirely different mechanism of growth. This series of investigations uncovered also many other incorrect claims about the growth of human population, claims creating an unnecessary and unhelpful system of misleading interpretations and explanations.

The key conclusions based on the discussions presented in all four publications, the three earlier publications (Nielsen aka Nurzynski, 2013a, 2013b, 2013c) and the current discussion, can be summarised as follows:

1. The concept of the Epoch of Malthusian Stagnation, the first stage of growth proposed by the Demographic Transition Theory is scientifically unacceptable. Many of its claims have to be either accepted by faith or are contradicted by empirical evidence.
2. Even large fluctuations in birth and death rates have no effect on the growth of human population.
3. Even sizable oscillations in the average difference between birth and death rates (growth rates) may only have a minor influence on the growth of human population.
4. Decreasing birth and/or death rates do not necessarily indicate a transition to a new stage of growth reflected in a noticeable change in the growth trajectory of human population.
5. The widening or narrowing gap between birth and death rates does not necessarily indicate a transition to a new stage of growth reflected in a noticeable change in the growth trajectory of human population.

6. Empirical evidence strongly suggests that the mechanism of Malthusian stagnation does not, and did not, work.

7. Empirical evidence strongly suggests that positive checks do not trigger the mechanism of Malthusian stagnation but that they activate the powerful and effective mechanism of Malthusian regeneration (Malthus, 1978; Nielsen aka Nurzynski, 2013b).

8. Positive checks do not suppress the growth of population but stimulate it.

9. Malthusian regeneration mechanism quickly repairs any damage caused by positive checks and allows for the growth trajectory to remain generally undisturbed.

10. There was only one event, a combination of five closely-spaced demographic catastrophes with the combined death toll of around 139 million, which appears to have created a small but noticeable disturbance in the growth of human population.

11. Positive checks could not have created the Epoch of Malthusian Stagnation.

12. High birth and death rates are not associated with small but with large growth rate.

13. Demographic catastrophes in the past were too weak to affect the growth of human population.

14. The growth of human population follows an exceptionally stable trajectory with no signs of random oscillations, distortions or irregular behaviour.

15. The Epoch of Malthusian Stagnation, the first stage of growth proposed by the Demographic Transition Theory, did not exist.

16. Random oscillations and distortions in the growth trajectory of human population cannot testify about the existence of Malthusian trap because they do not exist.

17. The Malthusian trap did not exist.

18. The escape from the Malthusian trap never happened. There was no escape because the trap did not exist.

The last conclusion invites further exploration. We would have to check whether there was any form of escape, because even if Malthusian trap did not exist, there might still have been some noticeable acceleration in the growth of human population that could be interpreted as an escape. This issue, which represents a part of the investigation aimed at a correct understanding of human population dynamics, is addressed in the forthcoming book mentioned in the footnote to the *Introduction*.